\documentclass[%
 reprint,
 aps,
 pra,
]{revtex4-2}

\usepackage{graphicx}
\usepackage{dcolumn}
\usepackage{bm}
\usepackage{makecell}
\usepackage{lineno}
\usepackage{float}
\usepackage[caption=false]{subfig}
\usepackage{comment}

\begin{document}

\preprint{APS/123-QED}

\title{Coulomb Expansion of Cold Non-Neutral Rubidium Plasma}

\author{Michael A. Viray}
\author{Stephanie A. Miller}
\altaffiliation{Present Address: Terumo Heart, Inc., Ann Arbor, MI 48103, USA}
\author{Georg Raithel}%
\affiliation{%
 Department of Physics, University of Michigan, Ann Arbor, Michigan 48109, USA
}%

\date{\today}

\begin{abstract}
We study the expansion of a cold, non-neutral ion plasma into the vacuum. The plasma is made from cold rubidium atoms in a magneto-optical trap (MOT) and is formed via ultraviolet photoionization. We employ time-delayed plasma extraction and imaging onto a position- and time-sensitive micro-channel plate detector to analyze the plasma.  We report on the formation and persistence of plasma shock shells, pair correlations in the plasma, and external-field-induced plasma focusing effects. We also develop trajectory and fluid descriptions to model the data and to gain further insight. The simulations verify the formation of shock shells and correlations, and allow us to model time- and position-dependent density, temperature, and Coulomb coupling parameter, $\Gamma({\bf{r}},t)$. This analysis both reaffirms the presence of shock shells and verifies that the experimental plasma is strongly coupled.

\end{abstract}

\maketitle


\section{\label{sec:intro}Introduction}

The development of atom trapping and cooling processes \cite{metcalf} has made it possible for researchers to photo-excite cold plasmas in the laboratory \cite{killian99, kulin00, pohl03, pohl04}. This enables models that scale to hard-to-access plasmas that occur, for instance, in astrophysical environments (insides of stars and gas planets) \cite{woolsey}, magnetic-confinement fusion \cite{mckenna, wessels18}, and inertial-confinement fusion \cite{djotyan18, jiang19}. Recent work in cold plasma physics has explored plasma laser cooling \cite{langin19}, pair correlations \cite{lyon17, murillo07, ott14, desbiens16}, dual-species ion collisions \cite{sprenkle19, boella16}, Rydberg atom-plasma interactions \cite{crockett18}, plasma field-sensing applications \cite{feldbaum02, anderson17, weller19}, and quenched randomness and localization \cite{sous19}.

One area of interest in laboratory plasma physics is the expansion of plasma into vacuum \cite{murphy14, morrison15, forest18}, a topic that has been of interest, in large parts, because of its parallels to astrophysical systems \cite{samir83, sack87}. The central topic of this paper is non-neutral-plasma expansion, also known as a Coulomb explosion \cite{feldbaum02, bychenkov05, grech11}.  Expanding non-neutral plasmas can exhibit several interesting phenomena such as vortices \cite{gould95} and Bernstein modes \cite{walsh18}. The phenomena of greatest interest in our paper are shock shells and strong coupling. Shock shells form in Coulomb explosions if, initially, the outer layer of ions is less dense than the center \cite{kaplan03, drake}. If so, intermediate layers of the plasma tend to catch up with peripheral layers, and form a higher-density shock shell. Shock fronts form in both spherically-symmetric and cylindrically-symmetric plasmas \cite{kaplan03}. They can also form in a wide variety of non-neutral plasmas, such as relativistic electron clouds \cite{zerbe18}.

Rapidly expanding plasmas are also intriguing thermodynamic systems \cite{forest18}. As the plasma expands, the temperature changes globally in time due to disorder-induced heating and adiabatic cooling \cite{pohl05}, and locally in space across the shock shells \cite{drake}. The temperature and density distributions expected for our plasma not only corroborate the presence of shock shells, but also lend themselves to a discussion of the plasma's Coulomb coupling strength, $\Gamma$, defined as

\begin{equation}
    \Gamma = \frac{q^2 / 4\pi \varepsilon_0 r_{WS}}{k_B T}
    \label{eq:gamma}
\end{equation}

Here, $q$ is the ion charge, $\varepsilon_0$ the vacuum permittivity, $k_B$ the Boltzmann constant, $T$ the plasma temperature, and $r_{WS}$ the Wigner-Seitz radius, which is related to the ion volume density $n_V$ by $r_{WS} = \sqrt[3]{3/(4 \pi n_V)}$. A plasma is considered strongly-coupled when $\Gamma \gtrsim 1$. While Eq.~\ref{eq:gamma} is often  employed in uniform plasmas in thermal equilibrium, it can also be used to describe plasmas that evolve in time or that are anisotropic, such as our expanding rubidium-ion plasmas, which are sufficiently close to local equilibrium. In this case, $\Gamma$ becomes both time- and position-dependent, based on local densities $n_V({\bf{r}},t)$ and temperatures $T({\bf{r}},t)$ defined on suitable spatial domains. While the determination of time- and position-dependent $n_V$, $T$ and $\Gamma$ is less straightforward than in steady-state systems,  
these parameters lend themselves to a more complete description of our expanding micro-plasmas.

In the present work, we study the Coulomb expansion 
of initially cold, laser-excited micro-plasmas prepared from a rubidium magneto-optic trap (MOT). The plasmas contain a few hundred ions and have initial diameters of tens of microns. Central to our work are ion imaging techniques using electric fields, and single-ion-counting micro-channel plate detectors. These methods have been used before on Rydberg atoms in order to observe blockade radii \cite{schwarz13}, van der Waals interactions \cite{thai15}, ionization spectra \cite{grimmel19, stecker19}  and tunnel ionization rates \cite{gawlas19}. Here, we use ion imaging to observe the free expansion, shock shells, coupling strength, and correlations in micro-plasmas as a function of expansion time. 

We model our experimental data with two types of computer simulations of plasma expansion, a molecular-dynamics model and a fluid model. These computer models serve two distinct purposes. First, they recreate the experimental two-dimensional (2D) imaging setup and verify the presence of shock shells in the lab plasma. Second, the computer models are used to analyze the 3D expansion of the plasma and to create time-dependent maps of $n_V$, $T$ and $\Gamma$. These maps of thermodynamic quantities provide further evidence for shock shells and shed light on the transient thermodynamic behavior of the system.

\section{Experimental Setup} \label{sec:setup}

We employ a two-MOT vacuum chamber with a primary MOT that contains a reservoir of trapped rubidium-87 atoms, and a secondary MOT from which the plasma is formed. Atoms are loaded from the primary MOT into the secondary with a pulsed pusher laser (1.5 mW peak power, 1 mm FWHM beam diameter, and 10 Hz repetition rate with duty cycle of 10\%). The secondary MOT is able to trap clouds of several $10^7$ \textsuperscript{87}Rb atoms at densities of up to $\sim 4 \times 10^{11}$~cm$^{-3}$. The region in which the laser-generated plasma is prepared is electric-field-zeroed to within 10~mV/cm using internal field compensation electrodes. The zeroing is based on observing the acceleration of small test ion samples as a function of the voltages on the compensation electrodes.

The experiment is run at a repetition rate of 10 Hz. In each cycle, a plasma is formed in the secondary MOT with a resonant two-photon photoionization process. Prior to plasma formation, the MOT light is turned off to prevent unwanted plasma outside the intended initial volume. A fraction of the cold atoms is resonantly driven by an 18 $\mu$s long, 780-nm laser pulse to the 5$P_{3/2}$ state. This laser beam has a Gaussian profile, with a waist $w_0$ of 9~$\mu$s, Rayleigh range of 326~$\mu$s, and central intensity of 105~$I_{\text{sat}}$ (the saturation intensity $I_{\text{sat}} = 1.6$~mW/cm$^2$). Five microseconds after the 780-nm pulse is turned on, the 5$P_{3/2}$ atoms are ionized with a 10 ns, 335-nm ultraviolet pulse from a Q-switched, frequency-tripled Nd:YAG laser. This wavelength is well above the ionization threshold wavelength of 479.1 nm for atoms in the 5$P_{3/2}$ state. Atoms in the intermediate state become ionized, and the liberated valence electrons have a kinetic energy of 0.9 eV, which is sufficient for the electrons to escape the plasma cloud. The initial temperature of the trapped atoms is roughly 100~$\mu$K, but the photoionization is accompanied with recoil heating to $\approx$44~mK, the initial temperature of our plasmas. While the 355-nm laser is linearly polarized, the plasmas we work with are dense enough that the laser polarization has no notable effect on ion recoil velocity distribution or initial temperature. The 355-nm pulse has a diameter of $\sim 2$~mm, which is much larger than the size of the 780-nm beam. Hence, the initial geometry of the plasma column is determined by the 780-nm beam size and its central intensity, and the MOT size. Further, the fluence of the 355-nm pulse is $\lesssim 10^{16}$~cm\textsuperscript{-2}, corresponding to a photoionization probability of the 5$P_{3/2}$ atoms of $\lesssim 10$ percent, leading to up to several hundred ions in the initial plasma volume.

Regarding the initial state of the plasma, it is also noted that the initial temperature of the trapped atoms of roughly 100~$\mu$K as well as radiation pressure from the 780-nm pulse are negligible. This is because the photo-ionization is accompanied by a fixed recoil energy of $5.8~\mu$eV (equivalent to an initial ion temperature of $\approx$44~mK), with an anisotropy given by the linear polarization of the 355-nm photo-ionization laser. Details of the initial ion velocity distribution (other than average kinetic energy) are not important, because the plasmas are dense enough to undergo efficient disorder-induced heating and thermalization on a time scale of only a few hundred ns.

\begin{figure}
    \centering
    \includegraphics[width = 8 cm]{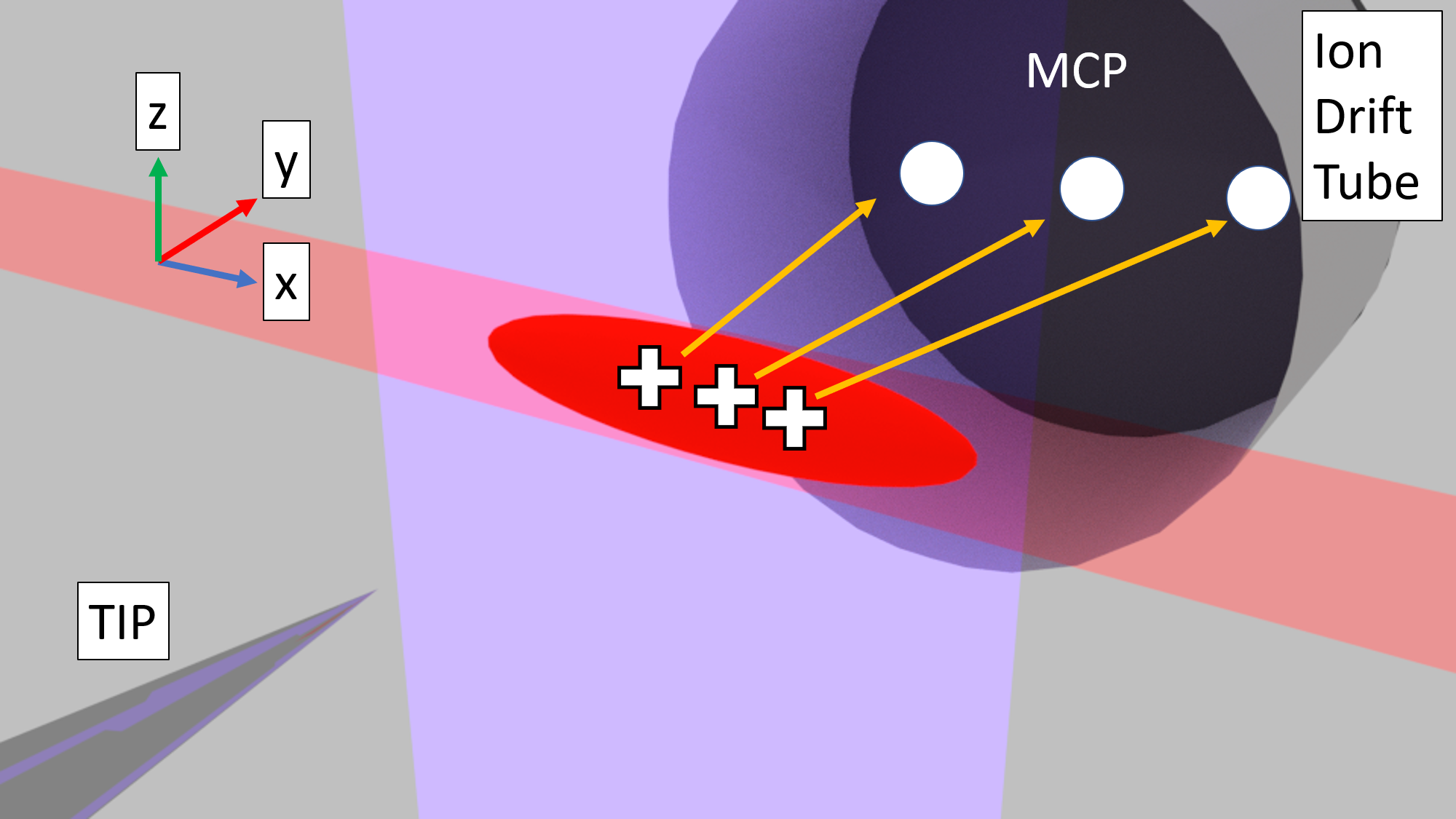}
    \caption{Rendering of experimental setup and process of plasma formation and imaging (not to scale). The positively charged rubidium ions are accelerated by the TIP electric field through the ion drift tube to the MCP, where they produce blips of a few hundred $ns$ decay time that are imaged with a camera.}
    \label{fig:config}
\end{figure}

The secondary MOT chamber contains a needle-shaped, beryllium-copper tip imaging probe (TIP), which is positioned $\sim 2$ mm away from the excitation region. Figure~\ref{fig:config} shows the orientation of the plasma, TIP, and other experimental components. After the plasma has been formed, it undergoes free expansion for a variable wait time $\tau$. After the wait time, a high voltage step pulse with a 90\% rise time of 76 ns is applied to the TIP, producing a divergent strong electric field. The ions are accelerated by this electric field towards a micro-channel plate (MCP) located about 25~cm away. Upon impact on the MCP, the ions produce bright spots on a phosphor screen which provide information on the plasma ion positions at the time of extraction. The MCP detection efficiency is 30-50\%, according to manufacturer's specifications. A CCD camera takes a picture of the phosphor screen for each experimental cycle. The single-ion resolving images represent the plasma ion distributions after the free-expansion time $\tau$, projected onto a plane transverse to the extraction trajectory. The magnification factor of the imaging setup can be varied by adjusting the distance between the TIP and the excitation region and can be calibrated by translating the excitation region by known distances. The spatial structure of the plasma transverse to the MCP plane (i.e. along the extraction trajectory) is analyzed by measuring its time-of-flight distribution to the MCP using a multichannel scaler (SRS Model SR430).

\section{Computer Modeling} \label{sec:model}

We use two approaches to model our system. In molecular-dynamics simulations, we calculate trajectories of up to 1000 ions, accounting for all external and inter-particle Coulomb forces. A numerically generated map of the electrostatic potential along the ion-imaging path between the TIP and the MCP is implemented to model the ion imaging. While this approach is computationally expensive due to the need to account for binary forces, it has the strengths that (1) it is exact, limited in particle number only by computing power, that (2) the ion-extraction timing, the analysis of the ion images on the MCP, the ion time-of-flight data, and the pair correlation functions closely follow the experimental procedure, and that (3) valuable insights on macroscopic plasma parameters such as expansion velocity, density, temperature, coupling parameter etc. can be obtained on a grid of spatial domains as a function of time. The trajectory simulation is exact, because it accounts for microscopic and macroscopic electric fields (here, there are no significant magnetic fields), initial ion velocities and their angular distribution, initial spatial ion distributions, and Coulomb collisions during free expansion, ion extraction and imaging. The trajectory simulations are well-suited to model correlations and strong-coupling effects, in addition to modeling the overall expansion and imaging dynamics.

In the second approach, the plasma is modeled as a collision-less medium (“fluid”) with a fixed charge/mass ratio that evolves under the influence of the macroscopic plasma electric field.  In this collision-free, zero-temperature approach, the plasma is numerically treated as a discrete set of thin, inter-penetrable cylindrical or spherical shells. The initial photon-recoil-induced ion motion is ignored. The initial charges on the shells are given by the initial shell radii, shell thicknesses, and the initial charge distribution, which depends on ion excitation parameters and atom densities (chosen similar to those used in the experiment). The discrete radii of the shells are then propagated using Newton's equations.  Gauss's law is employed to track the macroscopic electric field at the locations of the shells. The dynamical variables in this model are the shell radii and velocities, and the electric field. This model ignores the effects of the microscopic fields and collisions; it can therefore not be used to model inter-particle correlations. Also, in its present form the model does not allow for the inclusion of extraction and imaging electric fields that do not share the symmetry of the charge distribution.  Nevertheless, the fluid model is numerically inexpensive and suitable for the generation of plasma density maps versus time during the free-expansion time of the plasma. In the fluid model, shock fronts appear as singularities of the plasma density. Comparison between the fluid and the molecular-dynamics models then allows us to distinguish macroscopic from micro-field effects.  A detailed analysis based on these models is given in Sec.~\ref{sec:thermodynamics}.

\section{Experimental Results} \label{sec:shockshells}

In our experimental work, plasmas are excited from Rb atom clouds in a MOT, as described in Sec.~\ref{sec:setup}. The MOT contains $3.8 \times 10^7$ atoms at a density of $4.1 \times 10^{11}$ atoms/cm$^3$, as determined by shadow imaging. The spatial profile of the initial ion clouds is given by the geometry described in Sec.~\ref{sec:setup}. The actual initial number of ions detected within the fixed field of view is 100 to 150, based on ion count and MCP detection efficiency. The initial central ion density is $n \approx 10^{10}$ cm$^{-3}$, as determined from ion number and excitation geometry. We observe this plasma for $\tau = 0$, 2, 4, 6, 8, and 10~$\mu$s of free expansion time. Ion-density images, time-of-flight data and pair correlation functions are averaged over 5000 experimental realizations. The magnification factor of the imaging for this data set is 54 times, with a diameter of the field of view on the MCP of 12~mm. Density and initial ion kinetic energy from photo-ionization recoil indicate an initial Debye length of $\lambda_D \approx 1$ $\mu$m, more than a factor of ten less than the short-axes initial diameter of the ion cloud, and an initial Coulomb coupling parameter of $\Gamma \approx 50$ (following Eq.~\ref{eq:gamma}). Shortly after plasma generation, temperature and coupling parameter undergo a rapid increase and decrease, respectively, caused by disorder-induced heating. Several hundred nanoseconds into the Coulomb expansion, adiabatic cooling associated with the free Coulomb expansion leads to a recovery of these parameters. Our molecular-dynamics simulations provide considerable insight into these dynamics, as discussed in detail in Sec.~\ref{sec:thermodynamics}.

\begin{figure*}
    \centering
    \includegraphics[width=\textwidth]{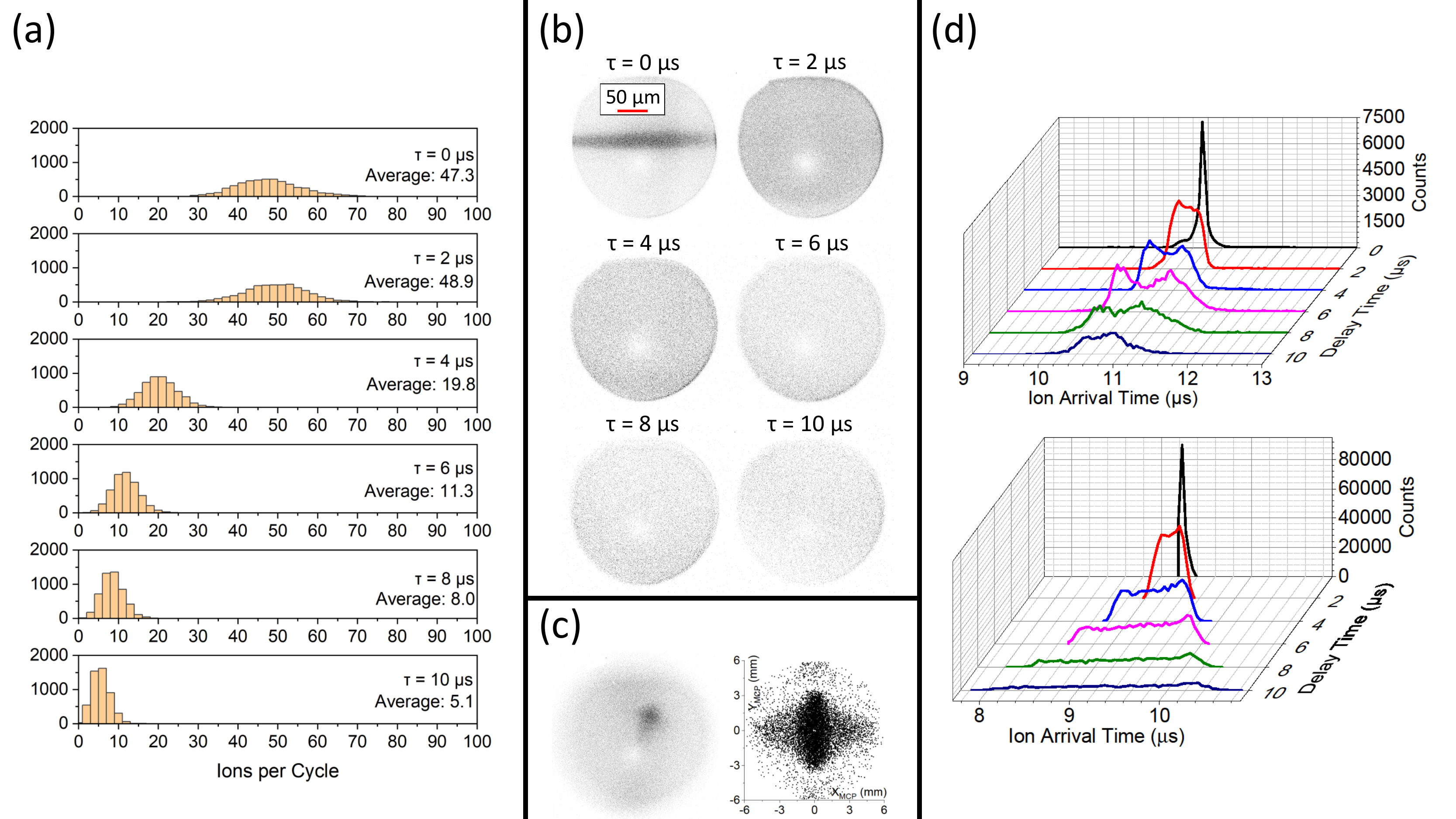}
    \caption{a) Histograms of ion counts per cycle for the indicated expansion times $\tau$. (b) Averaged pictures of the plasma images on the MCP plane for the indicated expansion times. The distance scale in the upper-left image corresponds to distance in the object plane. The light spot is an area of reduced ion detection rate; its origin is under investigation. (c) Experimental (left) and computational (right) MCP images of plasma at $\tau = 10$ $\mu$s in the presence of a quadrupole electric field. Instead of expanding outwards and leaving the field of view, the ions are refocused by the electric field and accumulate at a plasma focus, which maps onto a localized region on the MCP with well above-average count density. (d) Experimental (top) and computational (bottom) distributions of ion MCP arrival times (horizontal axis) for different plasma expansion times (tilted axis). To enhance visibility, the count axes for the experimental data are magnified by factors of 1, 2, 4, 8, 8, and 8, in ascending order of expansion times. Likewise, the count axes for the computational data are magnified by factors of 1, 8, 32, 64, 64, and 64.}
    \label{fig:shockshells}
\end{figure*}

Figures~\ref{fig:shockshells}~(a)~and~\ref{fig:shockshells}~(b) show the experimentally observed ion-count statistics and averaged images of plasma expansion. Starting from its initial cylindrical shape, seen at $\tau = 0$ $\mu$s, the plasma ions rapidly accelerate outward, leading to overall plasma expansion. At $\tau = 2$ $\mu$s, the plasma approximately fills the field of view along its initially short axis, with the average ion count still remaining approximately constant. For longer expansion times, the ions leave the field of view, and the average ion count monotonically decreases. Over the course of 10~$\mu$s of free expansion time, the ion count decreases to about 10$\%$ of the initial value.

Figure~\ref{fig:shockshells}~(d) shows experimental and simulated MCP arrival time distributions of the plasma ions for each expansion time $\tau$. The arrival time traces have been aligned on the time axis so that the firing of the TIP imaging voltage pulse is at $t=0$ (equivalent to the end of the expansion time $\tau$). As $\tau$ increases, the ion arrival time distribution develops a leading and a lagging peak, signaling the formation of shock shells. As the plasma expands radially, a fraction of ions move toward the TIP, while another fraction of ions move in the opposite direction toward the MCP. When the TIP electric field is engaged, the ions closer to the TIP experience a stronger electric field than ions that are farther away. Hence, the longitudinal position distribution of the ions maps onto a time-of-flight distribution in which shorter times of flight correspond to ions closer to the TIP at the onset of the TIP imaging voltage pulse. The peaks in the time-of-flight distribution correspond to shells of enhanced plasma density, or shock fronts. In our data in Fig.~\ref{fig:shockshells}~(d), the shock fronts begin to form at $\tau = 4$ $\mu$s and are most prominent at $\tau = 6$ $\mu$s. While this scenario is well-supported by the simulations that underly the numerical data in Fig.~\ref{fig:shockshells}, it is noted that experimental and simulated arrival time distributions begin to differ from each other at about $\tau = 6$ $\mu$s and onward. These differences are attributed to details of the ion-imaging electric field in the chamber that come to bear once the plasma has significantly expanded off-axis. These fields may, in part, be the result of higher-order multipoles of the TIP electric field, geometrical imperfections and stray potentials. These are not known and are therefore not included in the simulation.


Some experimental evidence of the importance of higher-order multipole electric fields is found in that it is not sufficient to zero the leading (dipolar) background electric field at the location of the initial plasma cloud. In addition to the homogeneous dipolar component, we also need to zero the linear (quadrupolar) components of the field in order to achieve relatively undistorted plasma images at long expansion times $\tau$. This is possible in our setup because it has nine electric-field compensation electrodes. Incomplete field-zeroing results in a quadrupole field that typically causes the expanding plasma to refocus along certain directions of space, while defocusing along other (orthogonal) directions. Figure~\ref{fig:shockshells}~(c) shows experimental and computational MCP images of a plasma at $\tau = 10$ $\mu$s in the presence of a quadrupole field that focuses the ions in the $xz$-plane and defocuses along the $y$-direction (coordinates defined in Fig.~\ref{fig:config}). In this case, the plasma focus generated by the quadrupole field manifests in a localized, kite-shaped region on the MCP of enhanced ion count density. The kite shape arises from astigmatism cause by the azimuthal dependence of the initial plasma cloud around the $y$-direction (see Fig.~\ref{fig:config}). In the present work, we have thoroughly field-zeroed the dipolar and quadrupolar field components to avoid this plasma focus effect. Nevertheless, remaining higher-order multipolar and other field perturbations are suspected as the main cause of deviations between experimental and simulated data at long expansion times.




As indicated above, the plasma may, in addition to shock fronts, also exhibit strong coupling, which should lead to structures in the pair correlation function $I(r)$ caused by Coulomb repulsion. From experimental data, we calculate pair correlations $I_{2D}(r)$ between ion counts in the images projected into the MCP plane. This is done for each expansion time, $\tau$. We first process the raw images with a peak finder algorithm, then obtain the pair correlation of the peaks, and then normalize the repetition-averaged pair correlation such that a value of one corresponds to an absence of correlations. Since the processing is performed on impact positions in the image plane (the MCP surface), which the ions reach $\sim 10~\mu$s after application of the high-voltage ion extraction pulse, the $I_{2D}(r)$ are a magnified, time-propagated representations of correlations that exist at the time instant of ion extraction. In the data displayed, projected distances on the MCP image plane are divided by magnification (here 54) to obtain projected distances $r$ in the object plane. For more details on this method, see \cite{schwarz13, thai15}.

\begin{figure}
    \centering
    \includegraphics[width = 8 cm]{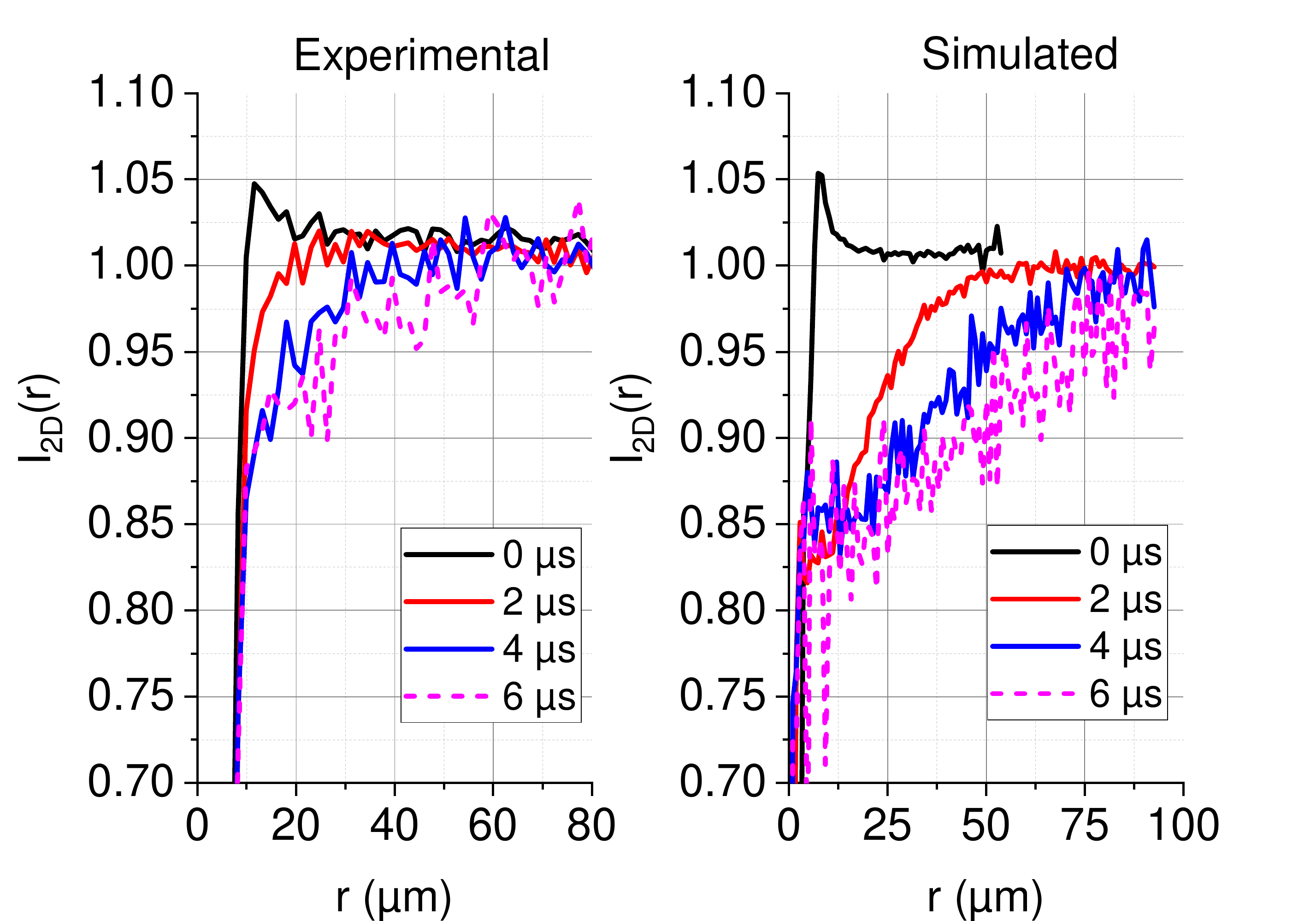}
    \caption{Projected, two-dimensional ion pair correlations $I_{2D}(r)$ for the indicated expansion times, measured vs radial coordinate in the $xz$ object plane. The left panel is for experimental and the right one for simulated data. Pair correlations for $\tau > 86$ $\mu$s are not shown because they are too noisy, as a result of the diminishing ion number in the images.}
    \label{fig:paircorrs}
\end{figure}

In Fig.~\ref{fig:paircorrs}, we show the functions $I_{2D}(r)$ in the object plane for the indicated values of $\tau$. The data are averages over 5000 repetitions for each expansion time. The $I_{2D}(r)$ reveal anti-correlation, $I_{2D}(r)<1$, out to a distance that increases with $\tau$, reaching $\approx 50~\mu$m at $\tau=6~\mu$s. At $\tau=2~\mu$s there is a region of enhanced correlation $I_{2D}(r)>1$, at $r \sim 10~\mu$m. These correlations are due to the micro-fields in the plasma, which cause disorder-induced heating~\cite{chen04} and subsequent buildup of strong correlations.

It is noted that the values of $I_{2D}(r)$ for small $r$ do not drop below $\approx 0.9$. This is due to the fact that the measured pair correlations represent 3D particle configurations projected into the $xz$-plane. Since the expanding plasma extends multiple correlation distances along the line of sight (the $y$-direction), the true 3D correlation function becomes washed out because distant, uncorrelated ions can have projections that appear close to each other, thereby mimicking an uncorrelated ion pair at a very short distance. Nevertheless, experimental and simulated data qualitatively agree in the degree of residual anti-correlation, $1-I_{2D} (r \sim 0) \approx 0.9$ in the experiment vs 0.85 in the simulation. Also, there a significant
degree of positive correlation, $I_{2D} (r \sim 0)-1 \approx 0.05$ at near-zero expansion time $\tau$, both in experiment and simulation. Further, the shapes of the functions $I_{2D}(r)$ and the range in expansion time $\tau$ over which the domain of anti-correlation expands qualitatively agrees between experiment and simulation. It is observed that the spatial extent of anti-correlation in the experiment is less than that in the simulation (see Fig.~\ref{fig:paircorrs}). We attribute this difference to the effects of higher-order multipolar fields and other field perturbations along the ion extraction trajectories.

The correlation function in three dimensions, $I_{3D}(r)$, would be a valuable observable because it does not suffer from the contrast reduction that comes with projection from 3D into 2D. While the experiment would require an expensive technology change to measure $I_{3D}(r)$, our molecular-dynamics simulations allow us to track $I_{3D}(r)$ during the free plasma expansion, as discussed in the following section.

\section{Detailed Numerical Study of Ion-Plasma Expansion} \label{sec:thermodynamics}

The experimental and computational work already presented provides evidence for ion correlations that may indicate strong coupling, as well as for shock fronts. In the following we study the expanding plasma by tracking the spatio-temporal evolution of a set of macroscopic parameters. Most of this computational work is done with a molecular-dynamics simulation that accounts for all Coulomb and external forces. Here we describe this simulation in some detail. We typically conduct 1000 to 10000 simulations for plasmas with initial ion phase-space coordinates that are randomly drawn from probability functions. The latter reflect the given initial macroscopic particle density and velocity distributions. Results for plasma density, temperature, etc. are averaged over the simulations.

Our particle system is large enough to describe its dynamics via time- and position-dependent macroscopic parameters in sub-volumes of the plasma that are close to a local equilibrium. To arrive at a suitable size and simple shape of the sub-volumes in the expanding plasma, we first restrict our study to a spherically symmetric system without external forces, so as to maintain macroscopic spherical symmetry. This allows us to break up the plasma into spherical partitions (shells) with a certain thickness, within which we find the localized thermodynamic parameters versus time.

As a guide for what constitutes a suitable shell number,
we note that the correlation length evident from the above discussed correlation functions is on the order of
$10\%$ of the typical system diameter. Based on this, the plasma is divided into ten partitions: the core and nine shells. Each partition contains a tenth of all plasma particles.
Further, for computation of the temperature and Coulomb coupling parameter we dynamically adjust the shell radii and thicknesses so that there always are $10\%$ of the ions in each of the ten shells. Since we run the simulation for ion numbers up to 1000, the shells contain fixed numbers of particles of up to 100. This means, for instance, that for a plasma of 150 particles the outer radius of the innermost shell, $i=1$, is defined as the average of the radial coordinates of the 15-th and the 16-th particle. Radial coordinates are measured relative to the symmetry center of the plasma. The dynamic variables in each simulation include the inner and outer radii of the shells, $r_{in,i}(t)$ and $r_{out,i}(t)$, with shell index $i=1,2, ..., 10$, and the ion-averaged shell radii, $r_i(t)$, and velocities $v_i(t)$. As the plasma expands, all shell radii expand. Ions are allowed to switch between adjacent shells so as to maintain a fixed number of ions in the shells.

In correspondence with the experiment, for the initial position density we choose (spherical) saturated Gaussians with radii of about $20~\mu$m. For the initial velocity distribution we use a fixed velocity magnitude given by the recoil that occurs in photo-ionization. Photo-ionization of the Rb 5P$_{3/2}$ state with 355-nm laser light has a kinetic-energy release of 0.9~eV, of which 5.8~$\mu$eV is picked up by the ions. The initial distribution of ion velocity angles is given by the linear polarization direction of the photo-ionization laser and spherical harmonics. Here, the plasma is dense enough that the velocity distribution locally thermalizes quickly enough so that details of the photo-ionization are negligible (only the initial velocity-magnitude has a small but measurable effect).


From $r_{in,i}(t)$ and $r_{out,i}(t)$ we directly get the time-dependent shell volumes. The ion densities $n_{V,i} (t)$ of the shells are then given by the fixed ion numbers in the shells divided by the shell volumes. To find temperatures and coupling parameters, it is important to first obtain the continuous macroscopic velocity function of the ions as a function of position. By symmetry, the macroscopic velocity function only has a radial component, $\bar{v}_r(r,t)$.
For each of the ten shells, $\bar{v}_r(r,t)$ is taken to be identical with a quadratic fit to the actual (microscopic) radial velocities of all ions in the shell. Since each of the shells $i$ has its own time-dependent velocity fit parameters,
we add an index $i$ to the velocity fit functions, $\bar{v}_{r,i} (r,t)$.
During the course of the up to 10000 individual simulations, the
accuracy of the fits $\bar{v}_{r,i}(r,t)$ is continually improved by taking the particle velocities of all earlier individual simulations into account when finding the fit functions $\bar{v}_{r,i} (r,t)$.
The temperature of the $i$-th shell, $T_{i}(t)$, is then given by the root-mean-square deviation of the particle velocities from the macroscopic average, which is given by the fit functions $\bar{v}_{r,i} (r,t) \hat{\bf{r}}$, with radial unit vector $\hat{\bf{r}}$. While the accuracy of the fits $\bar{v}_{r,i}(r,t)$ is critical in the described model, we note that a poor model for $\bar{v}_{r,i} (r,t)$ would result in higher temperatures and lower Coulomb coupling parameters. The results we show are understood to be close upper bounds for the actual macroscopic temperatures $T_{i}(t)$ and close lower bounds for the Coulomb coupling parameters, $\Gamma_i(t)$.

Once temperatures and densities are known, the simulation can readily calculate other macroscopic parameters such as the Debye length, the ion acoustic velocity, and the ion plasma frequency. Each of these parameters depends on shell index $i$ and time.

\begin{figure*}
    \centering
    \includegraphics[width=\textwidth]{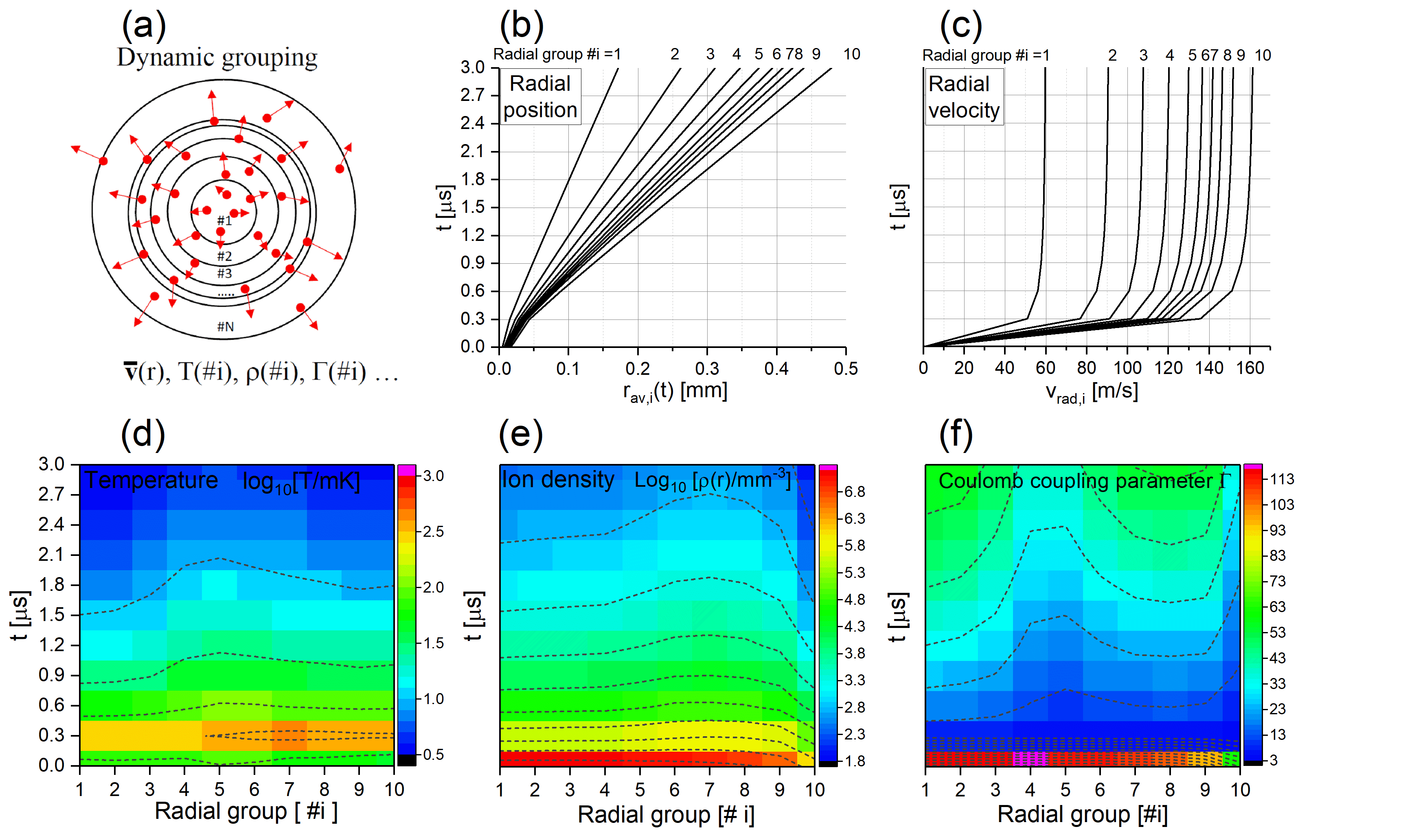}
    \caption{(a) Sketch of the radial-partition (shell) model explained in the text. The shell radii, the macroscopic velocity $\bar{v}_{r,i}(r,t)$ and other plasma parameters depend on shell index $i$ and time. (b) Average shell radii (horizontal axis) obtained for 125 ions with an initial cloud radius of about 20~$\mu$m vs expansion time (vertical axis). Shell indices listed on top. (c) Corresponding average shell velocities. (d) Ion density vs shell index (horizontal axis) and expansion time (vertical axis), displayed on a logarithmic color map. (e) Temperature vs shell index (horizontal axis) and time (vertical axis), displayed on a logarithmic color map. (f) Coulomb coupling parameter $\Gamma$ vs shell index (horizontal axis) and time (vertical axis), displaced on a linear color map.}
    \label{fig:th1}
\end{figure*}

In Fig.~\ref{fig:th1} we show density, temperature, and coupling parameter obtained from a set of simulations with 125 ions each. The depth of the space-charge potential well for electrons after plasma excitation is only 13~meV, {\sl{i.e.}} in our plasmas generated by photo-ionizing Rb 5P$_{3/2}$ atoms with 355-nm laser light all photoelectrons (energy 0.9~eV) near-instantaneously escape from the ion plasma and do not contribute to its dynamics. The system is, initially, very far away from any type of equilibrium, because the initial Coulomb potential energy of the ion cloud, after the electrons have escaped, exceeds the photo-ionization recoil energy of the ions by a factor near 1000. This implies that the initial kinetic energy of the ions does not play an important role, and that its initial Coulomb coupling parameter of near 100, computed from initial density and kinetic energy, does not have much physical meaning. We see from the temperature plot in Fig.~\ref{fig:th1} that the plasma as a whole heats up from tens of mK to near 1~K within a fraction of a microsecond due to disorder-induced heating. During that time the plasma reaches a dynamic steady-state, and temperatures and coupling parameters computed for the 10 radial partitions become physically meaningful. The ion plasma frequency drops from several MHz to about 0.5~MHz during that time, and the Debye length increases from sub-$\mu$m to several $\mu$m. In this phase, the coupling parameter $\Gamma$ traverses a low point of about six, {\sl{i.e.}} even in this hottest phase the plasma is strongly coupled. After about 0.5~$\mu$s, the plasma undergoes an adiabatic expansion phase of about one microsecond, during which the particles are still collisionally coupled while the overall acceleration caused by the Coulomb explosion fades and the spherical partitions approach a radially dependent terminal velocity. During this time, much of the initial plasma energy gets converted into the directed (non-thermal) kinetic energy of the expanding spherical partitions. This implies that during the adiabatic expansion the outer spherical partitions of the plasma turn highly supersonic, reaching Mach numbers of about 40 in the outermost partitions. The shock front develops, as seen in a relative rise in particle density in the outer radial partitions and a radial bunching of a few radial partitions around the partition $i=8$. Within the shock region, other macroscopic plasma parameters differ from their values in the intermediate radial region of the plasma. For instance, the temperature is about 20\% lower and the coupling parameter is about 30\% higher. Importantly, the coupling parameter increases, overall, and approaches $\sim 20$ at about 1~$\mu$s (within the shock region). Later into the expansion, the particles collisionally mostly de-couple, but the system continues to ballistically cool. After 3~$\mu$s of expansion time, the Coulomb coupling parameter exceeds 50, and the temperature has dropped to about 4~mK (from an initial photo-ionization recoil energy equivalent to 44~mK). Presently, we do not have an experimental method for direct verification of the simulated position- and time-dependent plasma temperatures.

The fluid model sketched in Sec.~\ref{sec:model} is well suited to qualitatively predict the formation of shock fronts.
In the fluid model, the plasma is split into radial or cylindrical shells, dependent on the symmetry of the problem. 
A shell in the fluid model is filled with an interpenetrable fluid with fixed charge-to-mass ratio and time-dependent and shell-dependent density. The fluid shells are propagated in the time-dependent macroscopic electric field of the system following Newton's equations of motion; the electric field is computed with Gauss's law. 
Thermal motion, particle collisions and micro-field effects are ignored. In the fluid model, the number of shells is chosen very large, so as to arrive at an accurate description of the fluid motion (here we use up to 10000). The shells in the fluid model are not to be confused with the partitions in the molecular-dynamics model. In the former, the shell positions and velocities themselves are propagated with equations of motion involving shell masses, charges and fields acting on the shells. Whereas, the latter merely serve as dynamic domains within which macroscopic plasma parameters of the ion system are computed. The ion mass points follow microscopic molecular-dynamics equations that have nothing to do with the partitions.

\begin{figure*}
    \centering
    \includegraphics[width = \linewidth]{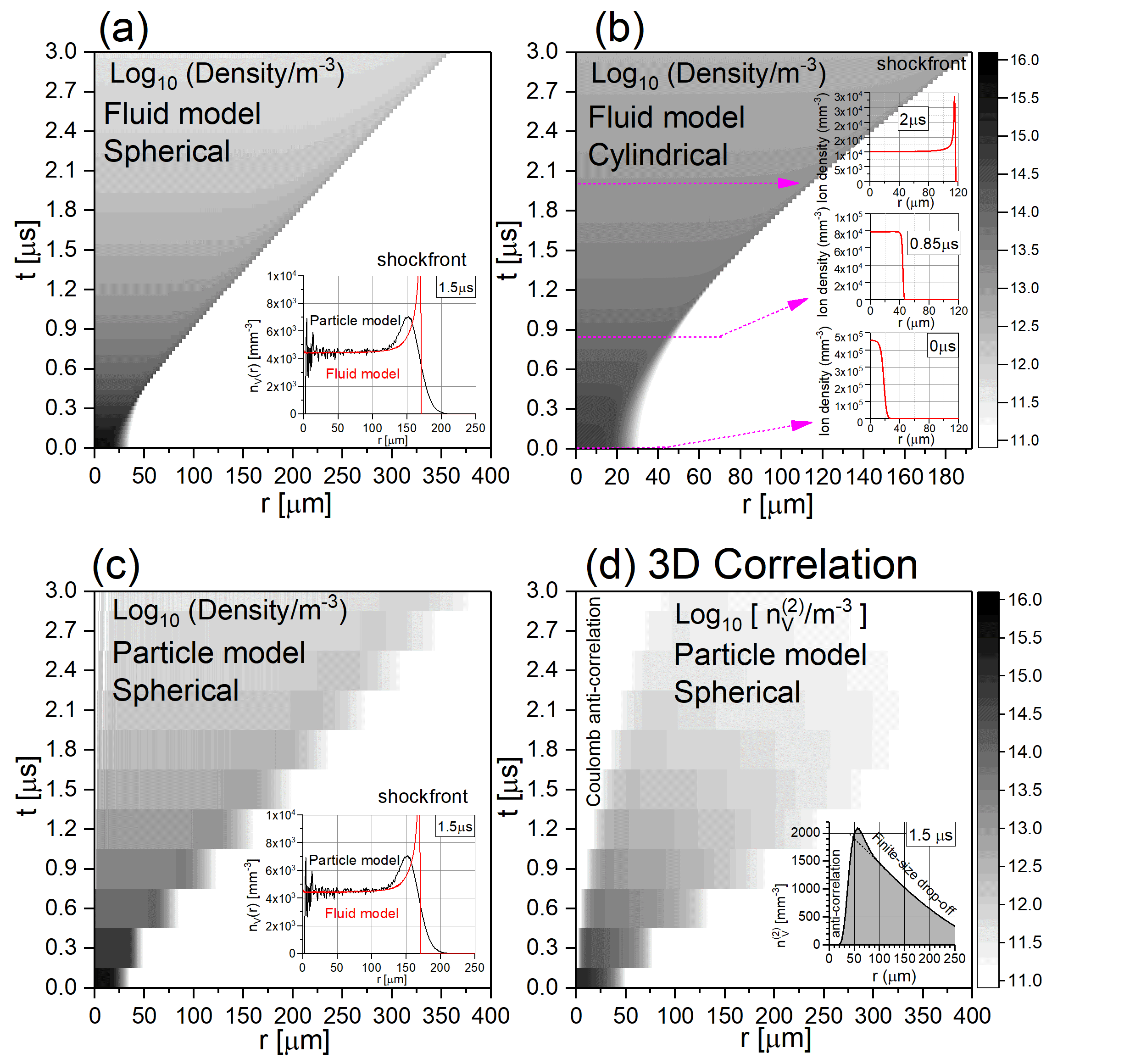}
    \caption{Fluid-model simulation of the expansion of an azimuthally symmetric plasma. Plasma density is shown vs radius (horizontal) and expansion time (vertical axis). A shock front develops at $0.85~\mu$s and becomes increasingly pronounced thereafter.}
    \label{fig:th2}
\end{figure*}

In Fig.~\ref{fig:th2}~(a) the shock front forms at about 0.3~$\mu$s expansion time and becomes more prominent later-on (see inset in Fig.~\ref{fig:th2}~(a)). In the fluid model, the onset of the shock front occurs when initially further-inside charged shells begin to overtake initially further-outside charged shells. The radial location of the shock front is given by the condition $d r_k (t)/ d k = 0$, where $r_k(t)$ is the radius of the $k$-th shell at time $t$, and $k$ is an integer counter that is assigned to the charged shells from the inside out (at time $t=0$). According to this equation, the singular behavior, which is equivalent to the shock front, marks a condition where a group of shells, with indices $k$ within a contiguous range, pile up at approximately the same radius, which is the shock-front location. This can only occur after some expansion time. At later times, the shock front becomes more pronounced, while in the interior region the plasma density becomes position-independent and continues to drop in time. In Fig.~\ref{fig:th2}~(c) and in the insets in Fig.~\ref{fig:th2}~(a) and (c) we compare the density of the fluid model with that of the particle model. It is seen that the overall behavior is similar, as expected. However, the shock front appears as a singularity in the fluid model, whereas it manifests as a moderate density enhancement in the particle model. This is due to the fact that fluid model fails to account for the granularity of the mass and charge distribution and micro-field effects, which all act to wash out the shock front.

In Fig.~\ref{fig:th2}~(b) we use the fluid model to describe the expansion of an cylindrically symmetric, infinitely long plasma with an initial radius of $\approx 20~\mu$m, an initial saturated Gaussian profile, and a total linear charge density of $5 \times 10^{5}$~e/m. These parameters correspond with those of the elongated plasma we have studied in our experiment.
It is seen in the simulations that the spherical and cylindrical systems undergo very similar dynamics, as expected. The insets in Fig.~\ref{fig:th2}~(b) highlight the differences in the radial density profiles before, at, and well after the shock front formation.

Finally, in Fig.~\ref{fig:th2}~(d) we use the particle model to obtain the three-dimensional pair correlation function in a freely expanding, spherically symmetric plasma. Due to the small finite size of the plasma, there is no well-defined radial distance beyond which the pair correlation function becomes stationary and equivalent with that of an uncorrelated system. Hence, a normalization akin to Fig.~\ref{fig:paircorrs} is not possible. Instead, we plot the density correlation function, $n^{(2)}_V (r)$, averaged over all particles in the plasma. (Since this quantity describes the average particle density at a three-dimensional distance $r$ from a randomly chosen ion, it has the dimension 1 over volume.) It is seen that the density correlation function rapidly develops an anti-correlated core within which $n^{(2)}_V (r)$ drops to near identical zero. The range of this core expands to beyond 50~$\mu$m at $3~\mu$s, and the rate of expansion keeps increasing, even after the particles collisionally decouple. This behavior accords with the finding that the Coulomb coupling continues to increase even at late times $\gtrsim 3~\mu$s. It may indicate that fine-scale rearrangement of particle positions, mediated by long-range Coulomb forces, continues to affect $n^{(2)}_V (r)$ and the Coulomb coupling, even after the system becomes largely non-collisional. 

Considering the relative clarity of the three-dimensional correlations in Fig.~\ref{fig:th2}~(d) in comparison with the two-dimensional ones shown in Fig.~\ref{fig:paircorrs}, it appears worthwhile to develop methods to measure the 3D correlation function. To reinforce this point, we note that a close look at Fig.~\ref{fig:th2}~(d) shows a significant ``overshoot'' immediately outside the anti-correlated core [compare $n_V^{(2)}(r)$ with the dashed line in the inset of Fig.~\ref{fig:th2}~(d)]. This may indicate short-range order. Similarly, the density function in Fig.~\ref{fig:th2}~(a) indicates a void region near $r=0$, as well as several quasi-periodic ripples. Future modeling and experiments on 3D correlation functions and structure factors may be necessary to elucidate these observations further.

\section{Conclusion} \label{sec:conclusion}

We have observed cold-ion plasma expansion by time-delayed ion extraction and ion imaging on a position-resolving particle detector. Results have been compared with two models: a particle model and a fluid model. From the experimental data, we have seen that as the plasma expands, shock fronts of high ion density form at the plasma-vacuum interface (the outer layers of the expanding plasma). Our computer simulations verify that these density build-ups are shock shells, which are accompanied by variations in other macroscopic plasma parameters (temperature, Coulomb coupling parameter). These results, as well as our experimental and simulated data on pair correlation functions, show the rich dynamics of micro-plasma explosions into the vacuum. In future work, we plan on observing non-neutral plasmas with different initial conditions (e.g. in initial density), investigating neutral plasmas, and applying atom-based electric-field metrology techniques to diagnose cold plasmas~\cite{anderson17, weller19}. The latter methods may enable a distinction between microscopic (Holtsmark) and macroscopic fields. To push the microscopic fields to higher values, amenable to experimental detection of plasma electric fields, one may use the plasma focus effect shown in Fig.~\ref{fig:shockshells}~(c) to take advantage of a transient spike in plasma density.



\begin{acknowledgments}
This work was supported by NSF Grant PHY-1707377. We thank David Anderson, Rydberg Technologies Inc, and Eric Paradis, Eastern Michigan University, for valuable discussions.
\end{acknowledgments}


\bibliography{plasmaexp}

\end{document}